\documentclass[12pt]{article}
 \pdfoutput=1

\usepackage{scicite}

\usepackage{times}

\usepackage{amsmath} 

\usepackage{graphicx}
\usepackage{braket}
\usepackage{siunitx}
\usepackage{booktabs}
\usepackage{multirow}

\newenvironment{sciabstract}{%
\begin{quote} \bf}
{\end{quote}}

\newcounter{lastnote}
\newenvironment{scilastnote}{%
\setcounter{lastnote}{\value{enumiv}}%
\addtocounter{lastnote}{+1}%
\begin{list}%
{\arabic{lastnote}.}
{\setlength{\leftmargin}{.22in}}
{\setlength{\labelsep}{.5em}}}
{\end{list}}

\title{Quantum control of photonic entanglement with a single sub-wavelength structure} 
\author
{Alexander B\"{u}se,$^{1,2\ast}$ Mathieu L. Juan,$^{1,2}$ Nora Tischler, $^{1,2}$\\ Vincenzo D'Ambrosio, $^{3}$ Fabio Sciarrino, $^{3}$ Lorenzo Marrucci, $^{4}$\\ Gabriel Molina-Terriza$^{1,2\ast}$\\
\\
\normalsize{$^{1}$Department of Physics \& Astronomy, Macquarie University, Sydney, Australia}\\
\normalsize{$^{2}$ARC Centre for Engineered Quantum Systems, Macquarie University, Sydney, Australia}\\
\normalsize{$^{3}$Dipartimento di Fisica,  Sapienza Università di Roma, Roma, Italy}\\
\normalsize{$^{4}$Dipartimento di Fisica, Universit\`a di Napoli Federico II, Napoli, Italy}\\
\\
\normalsize{$^\ast$To whom correspondence should be addressed;}\\
\normalsize{E-mail: alexander.buese@mailbox.org, gabriel.molina-terriza@mq.edu.au.}
}

\date{}

\begin{document} 




\maketitle


\begin{sciabstract}
  Quantum entanglement is the basic resource for most quantum in\-for\-ma\-tion schemes. A fundamental problem of using photonic states as carriers of quantum information is that they interact weakly with matter  and that the interaction volume is typically limited by the wavelength of light. The use of metallic structures in quantum plasmonics has the potential to alleviate these problems. Here, we present the first results showing that a single subwavelength plasmonic nanoaperture can controllably modify the quantum state of light. In particular, we experimentally demonstrate that two-photon entanglement can be either completely preserved or completely lost after the interaction with the nanoaperture solely depending on the relative phase between the quantum states. We achieve this effect by using a specially engineered two photon state to match the properties of the nanoaperture. The effect is fundamentally mediated by quantum interference which occurs at scales smaller than the wavelength of light. This connection between nano-photonics and quantum optics not only demonstrates an unprecedented control over light-matter interaction in the quantum limit, but also probes the fundamental limits of the phenomenon of quantum interference. 
\end{sciabstract}

\paragraph*{Introduction}
\label{introduction}

The progress in quantum information theory and the advances in the manipulation and control of quantum systems promise the advent of an era where quantum technologies will significantly change the fields of communication, computation and sensors. Even though some of these new technologies have already been successfully implemented, such as quantum key distribution \cite{Bennett.2014,Korzh.2015} and quantum clocks \cite{Katori.2011}, to fully develop the potential of these technologies we need to find new compact, reliable and robust quantum systems. In particular, we need to be able to control and process the elusive and delicate features of quantum entangled states, which are at the core of quantum technologies.

Light has become one essential ingredient in many of the quantum systems which may hold the key to the blossoming of these technologies. Quantum states of light are used as carriers of quantum information \cite{Northup.2014}, but classical fields are also used to control the quantum properties of atomic species \cite{Tan.2015}. This is due to the maturity of the sources of light (both classical and quantum) and a broad range of possibilities to control the features of light fields such as polarization, pulse duration, spatial modes, etc. Indeed, quantum states of light have been harnessed to demonstrate most of the building blocks needed for quantum information processing and more recently have been used in compact integrated waveguide arrays to perform simple quantum simulation tasks \cite{Owens.2011,Carolan.2015,Bentivegna.2015}.

Most of the achievements in quantum photonics are hindered by the same limitations as the classical processing of light: weak interactions with matter, which impede efficient nonlinear processes, and large devices with dimensions many times the wavelength of light. Plasmonic devices may hold the key to overcome these hurdles due to strong interaction with light, small volumes of interaction, and the possibility to engineer and fabricate suitable nanostructures to address particular tasks. Classical control of the plasmonic modes of nanostructures has already been achieved by the use of modulation of ultrashort pulses \cite{Aeschlimann.2007}, spatial control of the incident modes of light \cite{Volpe.2010}, and the use of different angular momentum modes \cite{ZambranaPuyalto.2014}. These achievements have allowed the processing of the classical properties of light at the nanoscale, the development of novel biosensors, and the enhancement of nonlinear processes for molecular characterization (See \cite{anker2008biosensing} for a Review on the subject).

In the context of quantum optics, plasmonic waveguides have shown that they can interact strongly with single-photon emitters and transport single photons \cite{Akimov.2007}. This property has already been used to implement plasmonic Hong-Ou-Mandel interferometers by combining waveguides of several micrometers length to form plasmonic beam-splitters \cite{Fakonas.2014}. In this way it has been shown that quantum correlations of propagating photons can survive the interaction with plasmonic structures. A similar effect was observed in large arrays of nanoapertures \cite{Altewischer.2002} where the phenomenon of extraordinary optical transmission was exploited \cite{Ebbesen.1998}. More recently, there have been a series of efforts in directly observing the quantum properties of the electronic coherent oscillations associated with plasmon resonances \cite{Tame.2013}. While quantum correlations and quantum entanglement have been observed to survive in structures with an overall size larger than the wavelength of light, the observation of the quantum properties of small metallic structures is paving the way to address the fundamental question whether photonic quantum entanglement can be processed or even survive the interaction with a single subwavelength structure. 

Here we present experimental evidence that symmetry protected quantum entangled photonic states can interact with single nanostructures without being affected. We also demonstrate that these nanostructures can very differently affect quantum entangled states which are only distinguished by their quantum phases. This effect, only explained through a novel quantum interference of the photonic states occurring at a subwavelength scale, constitutes a promising approach for the control of quantum states using nano-photonics.

%

\paragraph*{Nanoapertures as quantum photonic processors}

In this work we have focused our efforts on studying a very simple nanostructure consisting of an isolated circular aperture. This kind of structure is very versatile and has been used for nano-trapping experiments \cite{Juan.2009}, classical sensing of molecules \cite{Brolo.2012} and is essential in near field optical microscopy experiments \cite{Betzig.1992}. Our choice of the nanostructure was motivated by its high symmetry and the fact that it is well-studied. Even though there is no analytical solution of the Maxwell equations for this structure, many interesting properties have been found both theoretically and experimentally. In particular, nanoapertures mix the polarization components of an incident classical field, an effect that can be described as spin-orbit coupling \cite{Tischler.2014}. This mechanism was partially responsible for the loss of entanglement that photons experience when focused to a single nanostructure in an array \cite{Altewischer.2002}. One way of explaining this phenomenon is in terms of symmetries: The spin-orbit coupling results from the mixing of the helicities of the field while total angular momentum is conserved. This helicity mixing naturally occurs in non-dual structures, while the circular symmetry of the structure will impose the conservation of total angular momentum. A light field of well defined helicity can be decomposed in a series of plane waves having all the same circular polarization and can then take on only two values. As a consequence, a circular nanoaperture can simply be described as a beam-splitter where the modes are mixed in polarization -- the two helicities -- instead of being mixed into two different propagation directions (see Fig. 1). 


\begin{figure}[t]
	\begin{center}
		\includegraphics[width=0.9\textwidth]{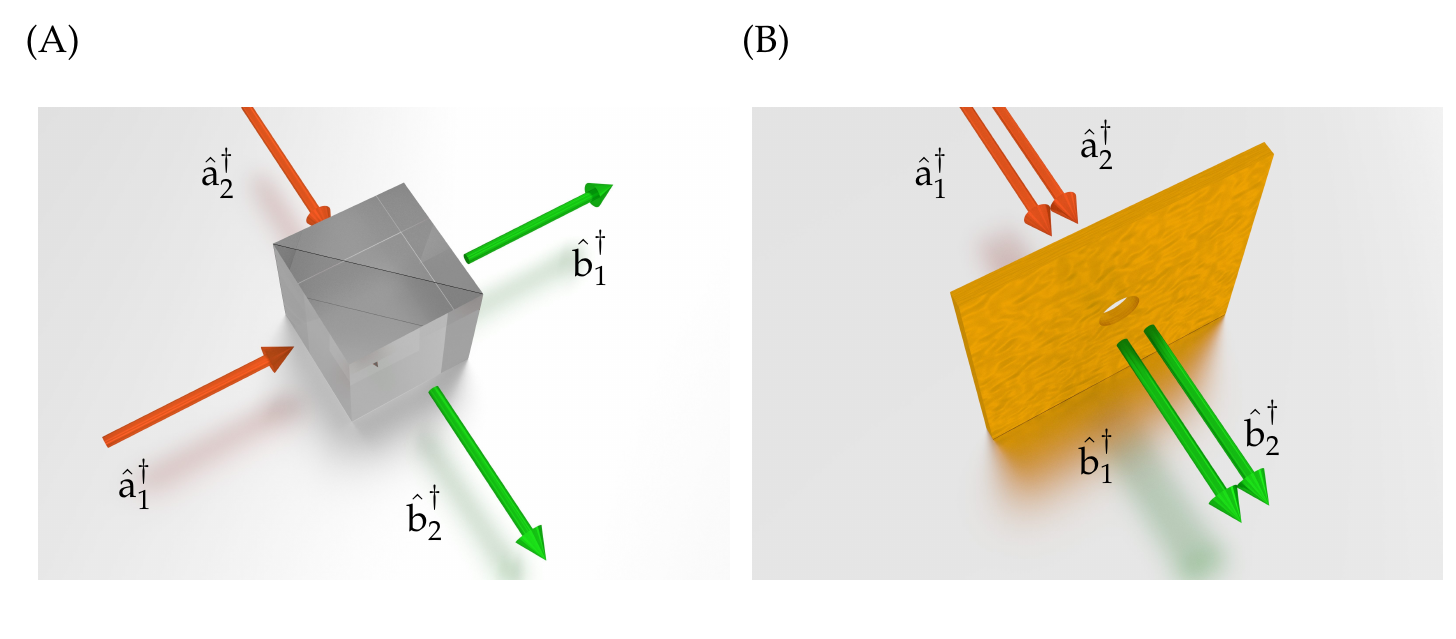}	
	\end{center}
	\caption{(A) Schematic of the input and output modes of a standard beam splitter, used for a Hong-Ou-Mandel type interference experiment. (B) In comparison a nanoaperture with two input and two output polarization modes.}
	\label{fig_comparison_standard_HOM_nano_aperture}
\end{figure}

Then, for a single photon $\hat{a}^\dag_{m,\Lambda}$ with total angular momentum $m$ and helicity $\Lambda$, the nanoaperture transforms the state in the following way:
	\begin{equation}
		\hat{a}^\dag_{m,\Lambda} \longrightarrow \alpha_{m,\Lambda} \hat{b}^\dag_{m,\Lambda} + \beta_{m,\Lambda} \hat{b}^\dag_{m,-\Lambda} \quad .
	\label{eq_nanoaperture_transform}
	\end{equation}
The probability for the helicity flip depends on the relative strengths of $\alpha$ and $\beta$, when passing through the nanoaperture. Note that the total angular momentum is conserved, since the aperture is cylindrically symmetric. Modes with the same total angular momentum and different helicities have distinct spatial mode structures, something that can be understood in the paraxial regime as a change in the orbital angular momentum. The amplitudes $\alpha$ and $\beta$ can be normalized if we only consider the light transmitted through the nanoaperture, as there will be significant losses due to light reflected from the metallic film, coupling to surface and localized plasmon modes and losses in the metal. For all these reasons the phases of the transmitted amplitudes are not locked as in the unitary beam-splitter. Nevertheless, the mirror symmetry of the structure imposes an important restriction on the amplitudes: $\alpha_{m,\Lambda}=\alpha_{-m,-\Lambda}$ and 
$\beta_{m,\Lambda}=\beta_{-m,-\Lambda}$. In particular, the subspace of modes with $m=0$ will transform onto itself: 
	\begin{align}
		\hat{a}^\dag_{0,+} &\longrightarrow \alpha \hat{b}^\dag_{0,+} + \beta \hat{b}^\dag_{0,-} \nonumber \\ 
		\hat{a}^\dag_{0,-} &\longrightarrow \alpha \hat{b}^\dag_{0,-} + \beta \hat{b}^\dag_{0,+} \quad ,
	\label{eq_nanoaperture_transform_tam_zero}
	\end{align}
where the sub-indices in the amplitudes have been omitted for simplicity. Considering two photons, there are three linearly independent basis states possible within this subspace:
	\begin{align}
		\ket{\Psi_0} &= \hat{a}^{\dagger}_{0,+}\hat{a}^{\dagger}_{0,-} \ket{0} \nonumber\\
		\ket{\Psi_+} &= \frac{1}{2}\left( \hat{a}^{\dagger}_{0,+}\hat{a}^{\dagger}_{0,+} + \hat{a}^{\dagger}_{0,-}\hat{a}^{\dagger}_{0,-} \right) \ket{0} \nonumber \\
		\ket{\Psi_-} &= \frac{1}{2}\left( \hat{a}^{\dagger}_{0,+}\hat{a}^{\dagger}_{0,+} - \hat{a}^{\dagger}_{0,-}\hat{a}^{\dagger}_{0,-} \right) \ket{0} \quad .
	\label{eq_symmetry_three_TAM0_states}
	\end{align}
Note that the states $\ket{\Psi_+}$ and $\ket{\Psi_-}$ are two photon entangled NOON states, which describe two-mode superpositions of N photons, where all the photons are in one of the two modes. Polarization two photon NOON states have already been used to beat the standard quantum limit of sensitivity in atomic spin species \cite{wolfgramm2013entanglement}. In this work, we show that further using the symmetries of these states, one can control their interaction with nanostructures in highly focused systems. 

The states $\ket{\Psi_0}$ and $\ket{\Psi_+}$ are mirror symmetric relative to any mirror plane containing the beam axis, while $\ket{\Psi_-}$ is the only mirror antisymmetric state within the whole subspace spanned by the three basis states.  Owing to the symmetry of $\ket{\Psi_-}$, the state is protected in this system. Even when both photons are transmitted through a circular aperture whose size is smaller than the wavelength, they should remain in this entangled state. This is clearly seen when applying the transformation (\ref{eq_nanoaperture_transform_tam_zero}) to the states $\ket{\Psi_-}$ and $\ket{\Psi_+}$:
	\begin{align}
		   	\ket{\Psi_-} & \longrightarrow \frac{1}{2} \left(\alpha^2 - \beta^2 \right)\left( \hat{b}^{\dagger}_{0,+}\hat{b}^{\dagger}_{0,+} - \hat{b}^{\dagger}_{0,-}\hat{b}^{\dagger}_{0,-} \right) \ket{0} \nonumber \\ 	
	 \ket{\Psi_+} & \longrightarrow \frac{1}{2} \left(\alpha^2 + \beta^2 \right)\left( \hat{b}^{\dagger}_{0,+}\hat{b}^{\dagger}_{0,+} + \hat{b}^{\dagger}_{0,-}\hat{b}^{\dagger}_{0,-} \right)\ket{0} + 2\alpha\beta \hat{b}^{\dagger}_{0,+}\hat{b}^{\dagger}_{0,-} \ket{0} \quad
	\label{plus_minus_transf}
	\end{align}
Most notably, while the only difference between the two input states resides in the phase between the two quantum states contributing to the entanglement, the difference after the subwavelength aperture is dramatic: one state survives structurally unaffected, while $\ket{\Psi_+}$ is mixed with $\ket{\Psi_0}$.

\paragraph*{Experimental realization}
\label{sec_experimental_realisation}

\begin{figure}[t]
	\begin{center}
		\includegraphics[width=0.9\textwidth]{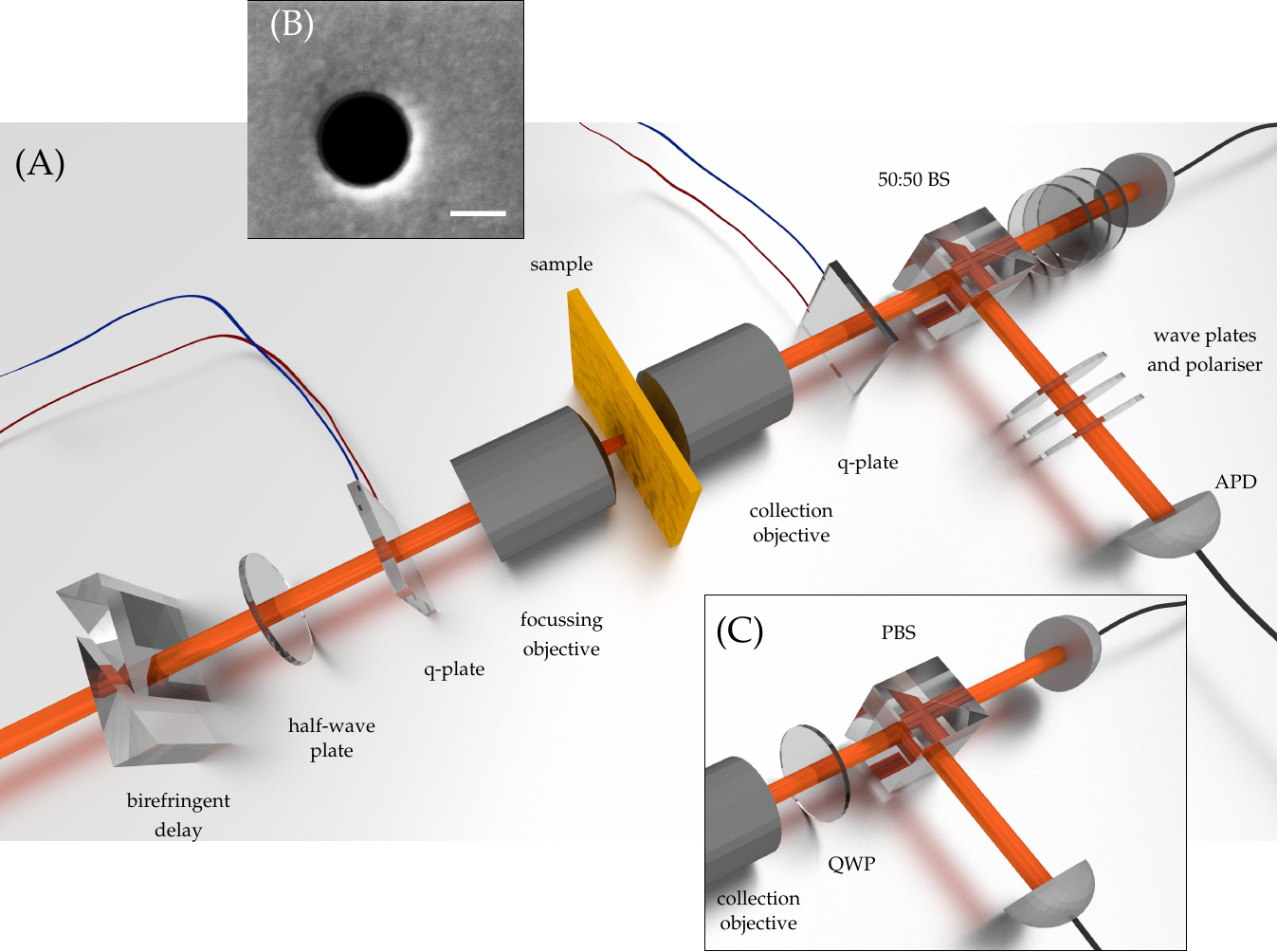}	
	\end{center}
	\caption{(A) Schematic of the experimental setup for transmission of protected quantum states through the nanoaperture. Starting from the left, the beam of orthogonally polarized photons passes through a birefringent delay line, a half-wave plate and a q-plate, which completes the state preparation. The beam is then focused onto the aperture (\SI{750}{nm} diameter, inset (B) shows a scanning electron microscope image of the nanoaperture iwth a scale bar of length \SI{500}{nm}) and re-collimated afterwards. After the transmission through the nanoaperture, a second q-plate reverses the transformation of the first one, removing the additional entanglement in orbital angular momentum. After that a 50:50 beam-splitter probabilistically separates the two photons and a set of waveplates and linear polarizers allows for correlation measurements in arbitrary polarization bases. Inset (C) shows a variation of the setup, where the quantum interference signature is measured directly. A quarter-wave plate and a polarizing beam-splitter project the output state onto the circularly polarized basis. Free-space avalanche photo detectors are used in both detection schemes. }
	\label{fig_tomography_setup}
\end{figure}

In order to study the two-photon state with and without the interaction with the nanoaperture we design an experimental set-up as depicted in Fig. \ref{fig_tomography_setup}. The two-photon generation is performed through spontaneous parametric down conversion in a collinear configuration (see supplementary material). The maximum Hong-Ou-Mandel interference visibility, attesting the two-photon indistinguishability, was found to be V = \SI{90}{\%} and is limited in our experiment by the frequency spectrum of the photons. The temporal overlap of the two generated photons was precisely controlled using a birefringent delay. A critical step in order to access the symmetry protected subspace for interaction with the nanoaperture is to transform the modes of the photons to those with total angular momentum zero. This step is achieved with a `q-plate'\cite{Marrucci.2006,Nagali.2009b} with $q=1/2$ which transforms our photons to the required modes. A half-wave plate placed before the q-plate allows us to select either the $\ket{\Psi_+}$ or the $\ket{\Psi_-}$ state (see supplementary material). The prepared two-photon state is then strongly focused with a microscope objective of numerical aperture NA$=0.85$ and subsequently collected with a second microscope objective with NA$=1.4$ to be finally analyzed.

\begin{figure}[htb]
	\begin{center}
		\includegraphics[width=0.8\textwidth]{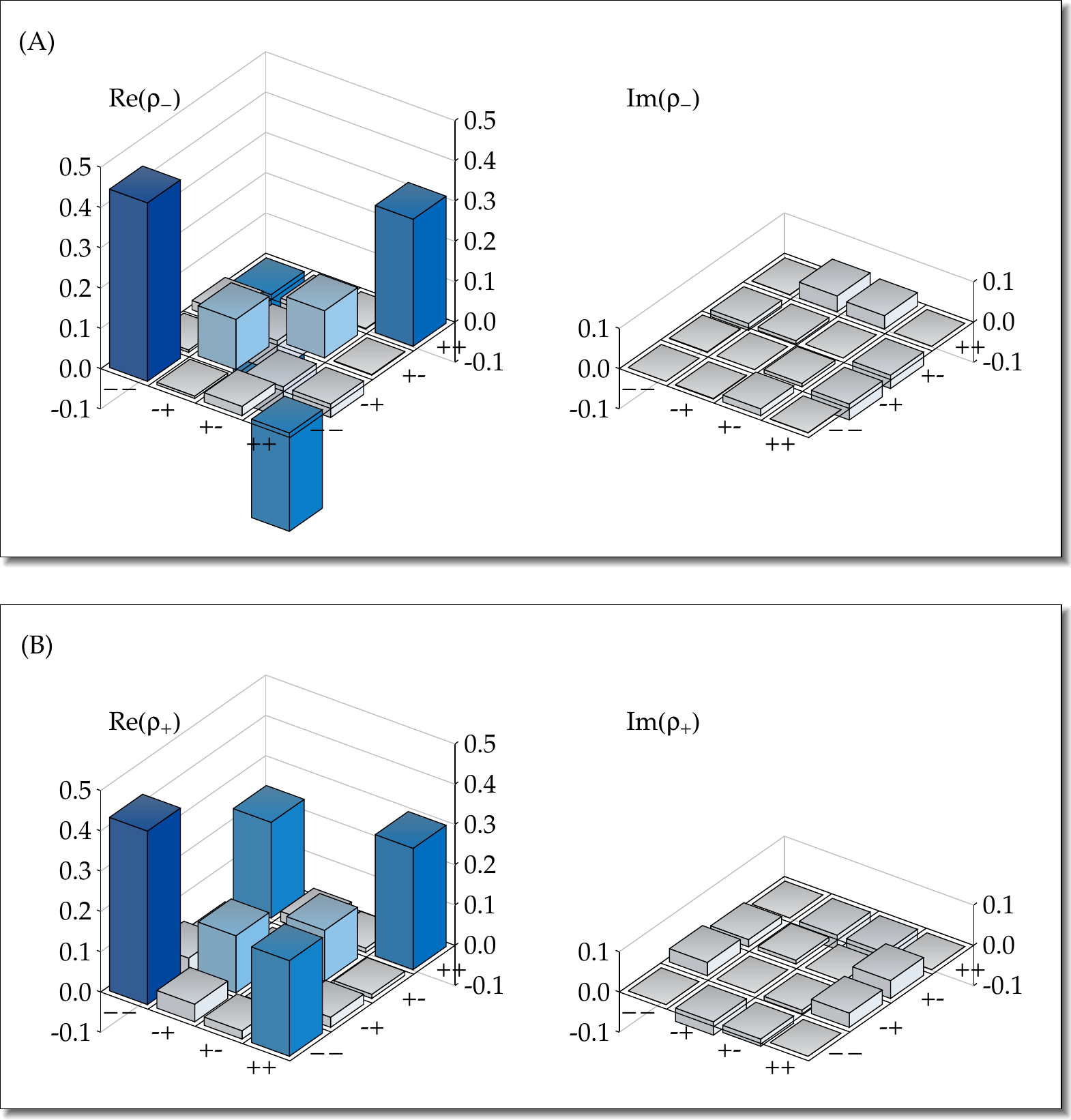}	
	\end{center}
	\caption{Reconstructed density matrices of the incident $\ket{\Psi_-}$ (A) and the $\ket{\Psi_+}$ (B) states before the interaction with the nanoaperture. The basis is given in terms of helicity. As expected from Equ. \ref{eq_symmetry_three_TAM0_states}, both $\ket{\Psi_-}$ and $\ket{\Psi_+}$ are a superposition of two-photon states with equal helicity. The only difference lies in the sign of the coherence as seen from the $++/--$ matrix element. Labels $+$ and $-$ in the axis, indicate the helicity state of the photons.}
	\label{fig_2016_03_03_M003_rho_LG_glass}
\end{figure}
\clearpage

In Fig. \ref{fig_2016_03_03_M003_rho_LG_glass} we present a tomographic reconstruction of the states in the absence of a nanoaperture in the system. As shown in Fig. \ref{fig_tomography_setup}, in order to tomographically analyze the photons we first transform them back to Gaussian modes with a q-plate reversing the effect of the initial one. In this way, the rest of the tomographical analysis is a simple polarization analysis as ideally the spatial modes of the photons are now identical. The reconstructed density matrices show a fidelity to the $\ket{\Psi_+}$ and the $\ket{\Psi_-}$ of more than $60\%$, limited by a weak incoherent contribution of the $\hat{a}^\dag_{+}\hat{a}^\dag_{-}\ket{0}$ modes. This loss of visibility is mainly due to the fact that we use bucket detectors to collect the photons, instead of projecting onto single mode fibers. The second q-plate, while providing the proper azimuthal transformation, does not readily project onto  the proper radial state, which together with slight misalignments and imperfections of the two q-plates, results in the creation of higher-order transverse modes. 

\begin{figure}[htb!]
	\begin{center}
		\includegraphics[width=0.8\textwidth]{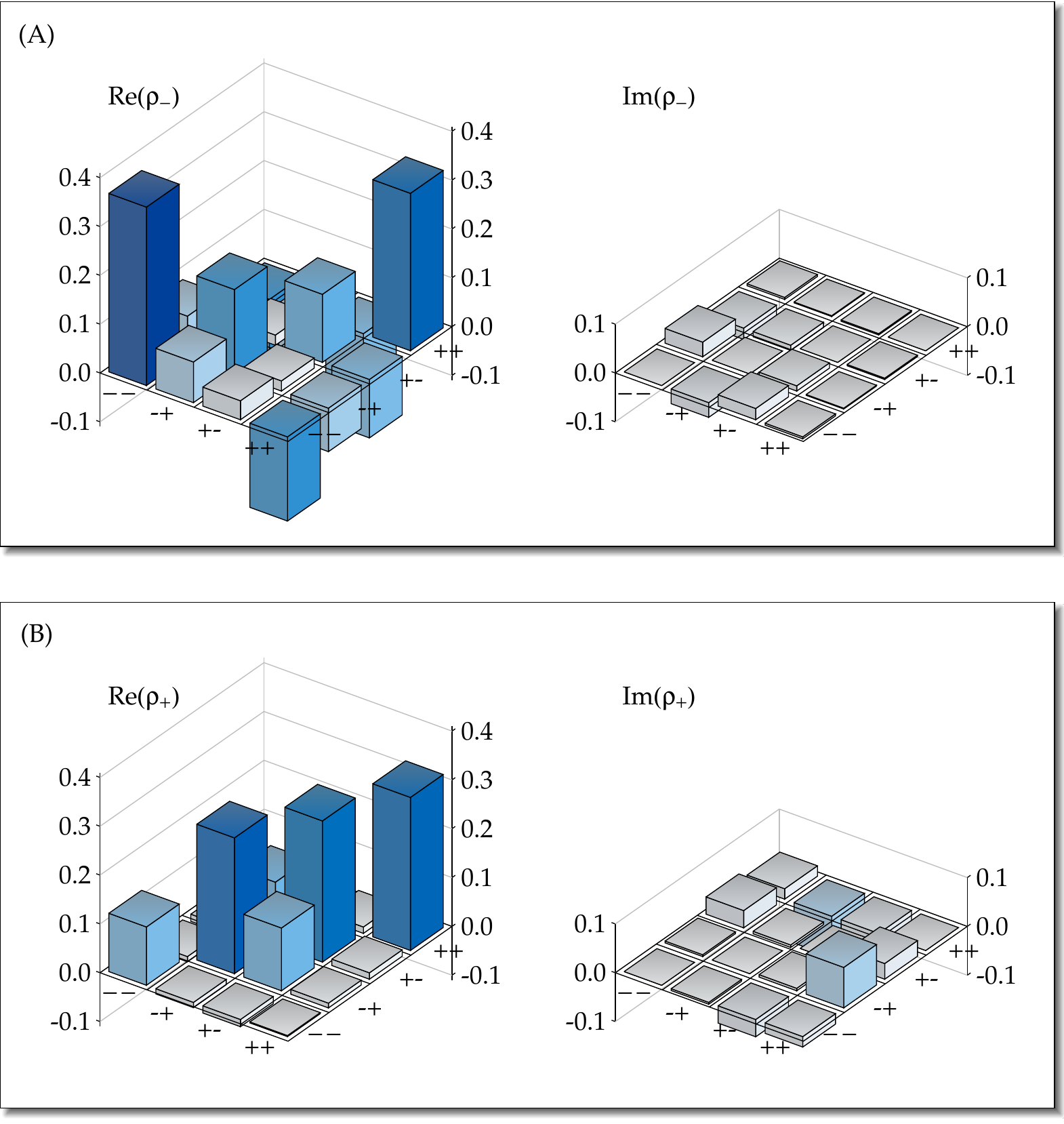}	
	\end{center}
	\caption{Reconstructed density matrices of the $\ket{\Psi_-}$ (A) and $\ket{\Psi_+}$ (B) states after interaction with the nanoaperture. In comparison to Fig. \ref{fig_2016_03_03_M003_rho_LG_glass}, the $\ket{\Psi_-}$ state is nearly unchanged, whereas $\ket{\Psi_+}$ was dramatically altered by the interaction with the nanohole. The appearance of terms with mixed helicity for the $\ket{\Psi_+}$ state is predicted by Equ. \ref{eq_symmetry_three_TAM0_states}.}
	\label{fig_2016_03_08_M013_rho_LG_H1}
\end{figure}

Once the two photons are allowed to interact with the nanoaperture, we perform another tomographic reconstruction of the states, as presented in Fig. \ref{fig_2016_03_08_M013_rho_LG_H1}. Now, the difference between the $\ket{\Psi_-}$ and $\ket{\Psi_+}$ states transformed through their interaction with the nanoaperture is remarkable. Where the minus state remains very similar to the state before the interaction (cf. Fig. \ref{fig_2016_03_03_M003_rho_LG_glass} (A)), with only slightly increased contributions from mixed polarization channels, the plus state (cf. Fig. \ref{fig_2016_03_03_M003_rho_LG_glass} (B)) has changed dramatically. Its coherence between the $\ket{--}$ and $\ket{++}$ contributions has completely vanished, while mixed terms are as strong as the pure $\ket{--}$ and $\ket{++}$ contributions. Interestingly, some coherence between the $\ket{-+}$ and $\ket{+-}$ terms has emerged. This coherence is consistent with a mixing with the $\ket{\Psi_0}$ state, rather than misalignments, while the extra noise in all polarization channels is mainly due to the decreased signal to noise ratio after the interaction. 

\begin{table}%
	\begin{tabular}{@{}lcccc@{}} \toprule
		 & \multicolumn{2}{c}{no interaction} & \multicolumn{2}{c}{interaction with aperture} \\ 
		 & $\ket{\Psi_-}$ & $\ket{\Psi_+}$ & $\ket{\Psi_-}$ & $\ket{\Psi_+}$ \\ \midrule
		concurrence	& $0.253\pm0.009$ & $0.233\pm0.009$	& $0.220\pm0.048$ & $0.020\pm0.019$		\\
		negativity &  $0.253\pm0.009$ & $0.230\pm0.009$	& $0.201\pm0.044$ & $0.017\pm0.016$		\\
		\multirow{2}{*}{fidelity to $\ket{--}
		\pm\ket{++}$}		& \multirow{2}{*}{$0.624\pm0.004$} & \multirow{2}{*}{$0.603\pm0.005$}	& \multirow{2}{*}{$0.515\pm0.024$} & \multirow{2}{*}{$0.270\pm0.020$}	 		\\ 
		&&&& \\ \bottomrule
	\end{tabular}
	\caption{Comparison of entanglement related quantities between minus and plus states before and after the interaction with the nanoaperture. The concurrence is an entanglement measure related to the entanglement of formation, where a value of zero indicates a completely mixed state and a value of one is given for a maximally entangled state. The same is true for the negativity, which is a measure for the entanglement that can be distilled from a given state. The fidelity gives the overlap to an ideal Bell state.}
	\label{table_entanglement_measures}
\end{table}

In table \ref{table_entanglement_measures} we present a summary of relevant quantities which quantify the extent of the interactions with the nanoaperture. We use two measures of entanglement: the concurrence \cite{Plenio.2007,Verstraete.2001}, related to the entanglement of formation, and the negativity, which is a measure for the entanglement that can be distilled from a given state \cite{Verstraete.2001}. In both cases, a value of zero indicates a completely mixed state and a value of one is given for a maximally entangled state. We find that both concurrence and negativity of the initial states indicate the presence of entanglement for both minus and plus states. However, although both states initially show the same degree of entanglement after the interaction only the minus state maintains its degree of entanglement, whereas the plus state's entanglement has virtually vanished. A very similar picture is given by the fidelity, which measures the state overlap with an ideal bell state. The fidelity is mostly preserved by the minus state through the interaction, whereas the fidelity of the plus state is greatly reduced, as is also obvious from directly comparing the density matrices in Fig. \ref{fig_2016_03_03_M003_rho_LG_glass} (B) and Fig. \ref{fig_2016_03_08_M013_rho_LG_H1} (B).

\paragraph*{Quantum interference in nanostructures}

\begin{figure}[ht]
	\begin{center}
		\includegraphics[width=0.9\textwidth]{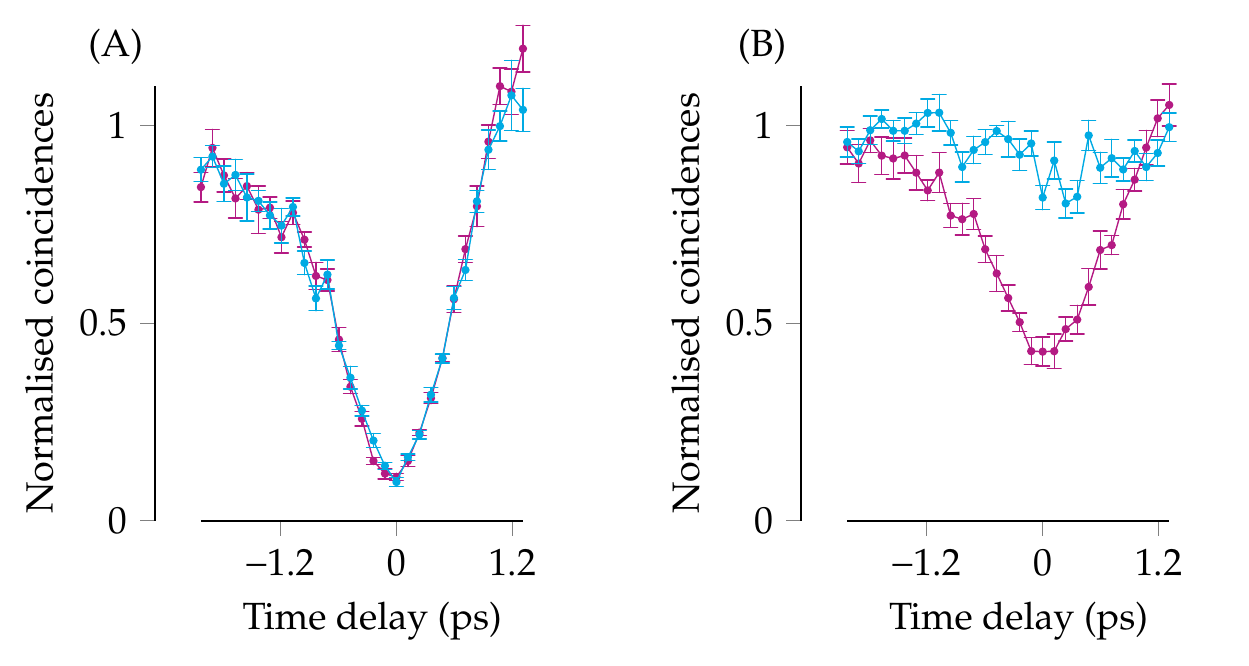}	
	\end{center}
	\caption{Normalised coincidence count rates for plus (blue) and minus (purple) states as a function of the time delay between the two photons. (A) Without going through the nanoaperture and (B) after transmission through the nanoaperture. Each coincidence rate has been measured ten times and the error bars span one standard deviation of the resulting distribution.}
	\label{fig_2016_01_22_M080_glass}
\end{figure}

The trademark of this dramatic difference between the transformation of the plus and minus states can be observed by replacing the tomographic reconstruction with simply projecting onto the circularly polarized states followed with Hong-Ou-Mandel type interference visibility measurements (see inset of Fig. \ref{fig_tomography_setup}). In Fig. \ref{fig_2016_01_22_M080_glass} (A) the effect of the delay between the two photons on the preparation of the states can be observed. At non-zero delay, the distinguishability of the photons prohibits the creation of the $\ket{\Psi_-}$ and $\ket{\Psi_+}$ states and raises the detection of coincidence events in the projected right and left circularly polarized channels (see Supplementary Material for a description of this effect). Only when the photons overlap in time (zero delay), either the plus or minus states are prepared and thus the coincidence in the cross polarized channels reach a minimum. When the photons do not interact with the nanoaperture, both states are conserved and consequently achieve the same visibility. On the other hand, the interaction with the nanoaperture distinguishes between these two entangled states and while the $\ket{\Psi_-}$ state retains a high visibility, the mirror symmetric state ($\ket{\Psi_+}$) completely loses its visibility, an effect which is consistent with a mixture with the other mirror symmetric state ($\ket{\Psi_0})$.

\begin{figure}[ht]
	\begin{center}
		\includegraphics[width=0.4\textwidth]{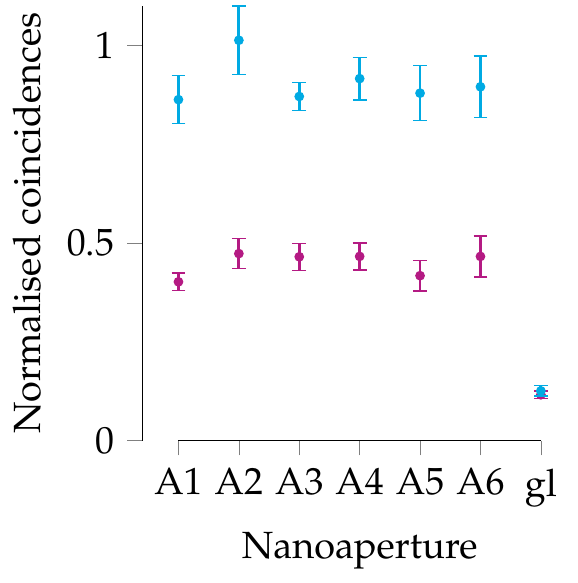}	
	\end{center}
	\caption{Quantum interference visibilities for a number of different apertures with nominally the same diameter. The coincidence rates have been measured at zero time delay. Each coincidence rate has been measured ten times and the error bars span one standard deviation of the resulting distribution.} 
	\label{fig_2016_01_22_largest_hole_dip_LG_oil}
\end{figure}

The observation of the normalized coincidence rates at zero delay for the plus and minus states thus provides a straightforward way of characterizing the behavior of the states under the interaction with the nanoaperture, and allows us to quickly and reliably analyze the interaction of these engineered states with a number of nanostuctures on one sample. For example, we compared nanoapertures with identical nominal size, proving that the stark difference between the plus and minus states is robust to the variations in the fabrication process of the nanostructures, as shown in Fig. \ref{fig_2016_01_22_largest_hole_dip_LG_oil}.

\paragraph*{Discussion}
\label{sec_entanglement_protection_discussion}

In this manuscript we have presented for the first time how engineered two-photon states can be processed through a single structure whose size is smaller than the wavelength of light. We have demonstrated that despite the extreme changes the electromagnetic modes undergo through strong focusing and the interaction with the nanostructure, by exploiting the symmetries of the nanostructure it is possible to engineer entangled states which are protected against decoherence and unwanted transformations. We have also demonstrated that this interaction strongly depends on the quantum phase between the entangled modes in such a way that a $\pi$ phase shift between the relative amplitudes of the states can be distinguished through the interaction with the nanostructures. From a fundamental point of view this is a result of the important impact this quantum phase has on the symmetry of the two photon state. Such a difference survives in the subwavelength regime and can be used to control the state using nanostructures.

This work also opens a promising approach to study the intricate interplay between plasmonic and quantum optics. Indeed, the behavior of the $\ket{\Psi_+}$ state and its deterioration through the interaction is rather puzzling. Looking at the density matrix of Fig. \ref{fig_2016_03_08_M013_rho_LG_H1} (B), we observe that coherence emerges for the $\ket{-+}$ term but not between the original state and the $\ket{+-}$ terms, nor does any coherence remain for the initial state, which is not directly obvious from equation \ref{eq_symmetry_three_TAM0_states}. One possible explanation of this phenomenon would be coupling to the localized or surface plasmon modes, which due to the monogamy property of entanglement \cite{Bru.1999} would give rise to decoherence between the $\ket{\Psi_0}$ and $\ket{\Psi_+}$ states.


\bibliography{Bibliography}

\begin{thebibliography}{10}

\bibitem{Bennett.2014}
C.~H. Bennett, G.~Brassard, {\it {Theoretical Computer Science}\/} {\bf 560,
  Part 1}, 7 (2014).

\bibitem{Korzh.2015}
B.~Korzh, {\it et~al.\/}, {\it {Nature Photon.}\/} {\bf 9}, 163 (2015).

\bibitem{Katori.2011}
H.~Katori, {\it {Nature Photon.}\/} {\bf 5}, 203 (2011).

\bibitem{Northup.2014}
T.~E. Northup, R.~Blatt, {\it {Nature Photon.}\/} {\bf 8}, 356 (2014).

\bibitem{Tan.2015}
T.~R. Tan, {\it et~al.\/}, {\it {Nature}\/} {\bf 528}, 380 (2015).

\bibitem{Owens.2011}
J.~O. Owens, {\it et~al.\/}, {\it {New J. Phys.}\/} {\bf 13}, 075003 (2011).

\bibitem{Carolan.2015}
J.~Carolan, {\it et~al.\/}, {\it {Science}\/} {\bf 349}, 711 (2015).

\bibitem{Bentivegna.2015}
M.~Bentivegna, {\it et~al.\/}, {\it {Science Advances}\/} {\bf 1} (2015).

\bibitem{Aeschlimann.2007}
M.~Aeschlimann, {\it et~al.\/}, {\it {Nature}\/} {\bf 446}, 301 (2007).

\bibitem{Volpe.2010}
G.~Volpe, G.~Molina-Terriza, R.~Quidant, {\it {Phys. Rev. Lett.}\/} {\bf 105},
  216802 (2010).

\bibitem{ZambranaPuyalto.2014}
X.~Zambrana-Puyalto, X.~Vidal, G.~Molina-Terriza, {\it {Nat. Comms.}\/} {\bf 5}
  (2014).

\bibitem{anker2008biosensing}
J.~N. Anker, {\it et~al.\/}, {\it Nature materials\/} {\bf 7}, 442 (2008).

\bibitem{Akimov.2007}
A.~V. Akimov, {\it et~al.\/}, {\it {Nature}\/} {\bf 450}, 402 (2007).

\bibitem{Fakonas.2014}
J.~S. Fakonas, H.~Lee, Y.~A. Kelaita, H.~A. Atwater, {\it {Nature Photon.}\/}
  {\bf 8}, 317 (2014).

\bibitem{Altewischer.2002}
E.~Altewischer, M.~P. {van~Exter}, J.~P. Woerdman, {\it {Nature}\/} {\bf 418},
  304 (2002).

\bibitem{Ebbesen.1998}
T.~W. Ebbesen, H.~J. Lezec, H.~F. Ghaemi, T.~Thio, P.~A. Wolff, {\it
  {Nature}\/} {\bf 391}, 667 (1998).

\bibitem{Tame.2013}
M.~S. Tame, {\it et~al.\/}, {\it {Nature Phys.}\/} {\bf 9}, 329 (2013).

\bibitem{Juan.2009}
M.~L. Juan, R.~Gordon, Y.~Pang, F.~Eftekhari, R.~Quidant, {\it {Nature
  Phys.}\/} {\bf 5}, 915 (2009).

\bibitem{Brolo.2012}
A.~G. Brolo, {\it {Nature Photon.}\/} {\bf 6}, 709 (2012).

\bibitem{Betzig.1992}
E.~Betzig, J.~K. Trautman, {\it {Science}\/} {\bf 257}, 189 (1992).

\bibitem{Tischler.2014}
N.~Tischler, {\it et~al.\/}, {\it {Light: Science {\&} Applications}\/} {\bf
  3}, e183 (2014).

\bibitem{wolfgramm2013entanglement}
F.~Wolfgramm, C.~Vitelli, F.~A. Beduini, N.~Godbout, M.~W. Mitchell, {\it
  Nature Photonics\/} {\bf 7}, 28 (2013).

\bibitem{Marrucci.2006}
L.~Marrucci, C.~Manzo, D.~Paparo, {\it {Phys. Rev. Lett.}\/} {\bf 96}, 163905
  (2006).

\bibitem{Nagali.2009b}
E.~Nagali, {\it et~al.\/}, {\it {Phys. Rev. Lett.}\/} {\bf 103}, 013601 (2009).

\bibitem{Plenio.2007}
M.~B. Plenio, S.~Virmani, {\it {Quantum Info. Comput.}\/} {\bf 7}, 1 (2007).

\bibitem{Verstraete.2001}
F.~Verstraete, K.~Audenaert, J.~Dehaene, B.~{De~Moor}, {\it {Journal of Physics
  A: Mathematical and General}\/} {\bf 34}, 10327 (2001).

\bibitem{Bru.1999}
D.~Bru{\ss}, {\it {Phys. Rev. A}\/} {\bf 60}, 4344 (1999).

\end{thebibliography}

\bibliographystyle{Science}


\begin{scilastnote}
\item This work was funded by the Australian Research Council's Centres of Excellence for Engineered Quantum Systems (EQuS). G.M.-T. also holds an Australian Research Council Future Fellowship.
\end{scilastnote}


\end{document}